\documentclass[runningheads]{llncs}
\usepackage[utf8]{inputenc} 
\usepackage[T1]{fontenc}    
\usepackage[colorlinks]{hyperref}       
\usepackage{url}            
\usepackage{booktabs}       
\usepackage{amsfonts}       
\usepackage{nicefrac}       
\usepackage{microtype}      
\usepackage{graphicx}
\usepackage{natbib}
\usepackage{doi}
\usepackage{graphicx}
\usepackage{svg}
\usepackage{rotating}
\usepackage{times}
\usepackage{todonotes}

\newcommand{\ignore}[1]{}

\title{To whom did my vote go? \thanks{This work was supported by the Australian Research Council (Discovery Project DP220101012, OPTIMA ITTC IC200100009) and by the European Union (ERC, CertiFOX, 101122653). Views and opinions expressed are however those of the authors only and do not necessarily reflect those of the European Union or the European Research Council. Neither the European Union nor the granting authority can be held responsible for them.  Andrew Conway is a member of the Secular Party; this paper has nothing to do with the Secular Party.}}

\author{
Andrew Conway\inst{1}
\and
Michelle Blom\inst{2}
\and
Alexander Ek\inst{3}
\and
Peter J. Stuckey\inst{4}
\and
Vanessa J. Teague\inst{5}
\and
Damjan Vukcevic\inst{6}
}
\authorrunning{Conway, Blom, Ek, Stuckey, Teague, and Vukcevic}

\institute{
\url{https://www.andrewconway.org}, Melbourne, Australia
\and
\email{michelle.blom@unimelb.edu.au}, School of Computing and Information Systems, University of Melboure,
Parkville, Australia
\and
Department of Computer Science, KU Leuven, Heverlee, Belgium
\and
Department of Data Science and AI, Monash University, Clayton, Australia
\and
Australian National University and
Thinking Cybersecurity Pty Ltd. 
\and
Department of Econometrics and Business Statistics, Monash University, Clayton,
Australia
}

\hypersetup{
pdftitle={To whom did my vote go?},
pdfauthor={Conway, Blom, Ek, Stuckey, Teague, and Vukcevic},
pdfkeywords={STV, Elections},
}

\begin{document}
\maketitle

\begin{abstract}
 Single Transferable Vote (STV) counting, used in several jurisdictions in Australia, is a system for choosing multiple election winners
 given voters' preferences among candidates. The system is complex and it is not always obvious how an individual's vote contributes to candidates' tallies across rounds of tabulation. This short paper presents a demonstration system that allows voters to enter an example vote in a past Australian STV election, and see: (i)~how that vote would have been transferred between candidates; and (ii)~how much that vote would have contributed to the tallies of relevant candidates, across rounds of tabulation.
\end{abstract}

\section*{The single transferable vote (STV)}

\enlargethispage{\baselineskip}

STV is a preferential multi-winner election system used widely in Australia (at the national, state, and local level), in Ireland and Malta (EU, national, and local level), and in New Zealand, Scotland, and the United States of America.  
In Australia, it is used to elect senators to the Australian Senate, and elect representatives in many state, territory and local council elections. Voters are required to express a preference ordering over (at least) some minimum number of parties or candidates. The tabulation process proceeds as follows \citep{conway2024idiosyncratic}:

\begin{itemize}
    \item Step 1: Each vote gets distributed to the highest-ranked candidate.
    \item Step 2: Calculate the value of a `quota', typically being slightly more than the number of valid votes
                  divided by one more than the number of vacancies (positions to fill). The idea is that it should
                  be impossible for more candidates than the number of vacancies to simultaneously reach a quota.

    \item Step 3: If a candidate has at least a quota of votes, they are considered elected.
    \item Step 4: If enough candidates are elected, stop.
    \item Step 5: If the number of vacancies not already filled equals the number of \emph{continuing} candidates
                  (those neither elected nor excluded), declare them elected and stop.
    \item Step 6: If an elected candidate had more than a quota number of votes, distribute the surplus votes (those over the quota)
                  to the next continuing candidate (neither elected nor excluded), and go back to step 3.
    \item Step 7: Exclude the candidate with the fewest votes. Distribute their votes
                  to the first continuing candidate on the preference list. Go to step 3.
\end{itemize}

The way STV is implemented across Australia varies. Mostly, these differences arise in Step 6, with different approaches used to re-weight the value of votes so that only a surplus-worth of votes are distributed to remaining candidates. 

\section*{A vote's journey}
Given the complexity of the STV counting algorithm, and the ballot papers---Australian STV elections can sometimes have hundreds of candidates standing for election---it can be hard for voters to get a sense of how their vote has contributed to the winners and losers of an election. 
We have developed a demonstration system that allows voters to explore the journey their vote took in any of the recent past Australian Federal Senate elections. This system is available at:
\url{https://vote.andrewconway.org}.

\begin{sidewaysfigure}
    \includegraphics[width=\textwidth]{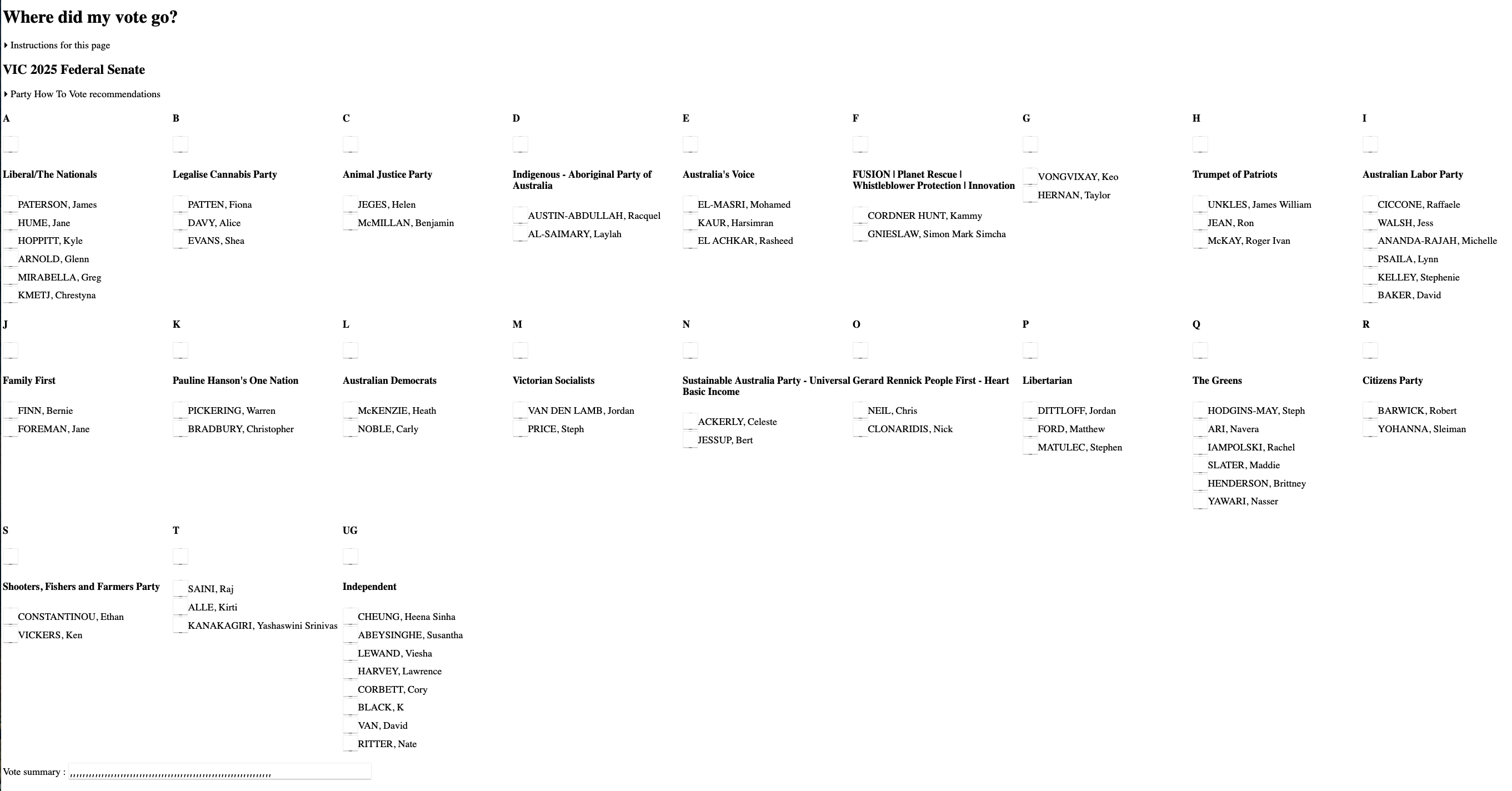}
    \caption{Ballot entry screen for the 2025 Federal Senate election for Victoria.}
    \label{fig:BallotEntry}
\end{sidewaysfigure}

Under `Currently available elections' for the `Federal Senate', a voter can select the year and state/territory of interest, which will take them to a page that summarises the results of that election. At the bottom of that page, under the title `Vote Search', the second link `see who your vote ended up help get elected' takes the voter to a page that recreates the ballot paper they would have filled out in that election. For the 2025 Federal Senate election for Victoria, \autoref{fig:BallotEntry} shows the ballot entry portion of the web page.

\begin{figure}[t]
    \includegraphics[width=\textwidth]{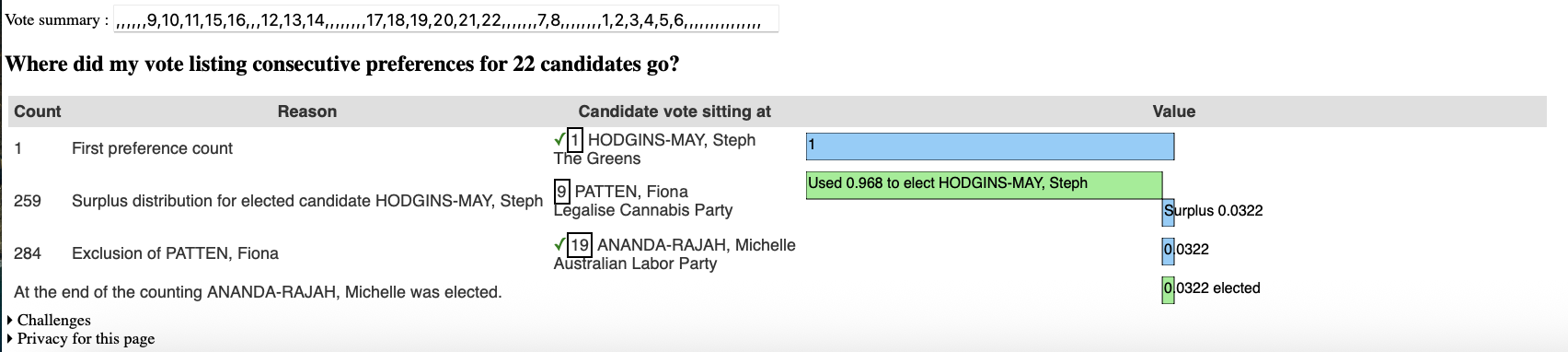}
    \caption{Vote journey for a ballot filled out according to The Green's How to Vote card in the 2025 Federal Senate election for Victoria.}
    \label{fig:ExampleVICResult}
\end{figure}

Using the links in the `Party How to Vote recommendations' section (click to open it), users can pre-fill the boxes associated with each group, party, and/or candidate. Alternatively, users can enter a custom vote. As the ballot is filled, the bottom of the screen will update to show the journey of that vote.  \autoref{fig:ExampleVICResult} shows the journey that a ballot would have taken if filled out according to The Green's How to Vote card.

The Green's How to Vote card ranked candidates from the following parties, in the stated order: The Greens (1--6); Victorian Socialists (7--8); Legalise Cannabis Party, LCP (9--11); Australia's Voice (12--14);  Animal Justice Party, AJP (15--16); and the Australian Labor Party, ALP (17--22).  \autoref{fig:ExampleVICResult} shows that a vote of this type starts with  Steph HODGINS-MAY, who is the first candidate in the Green's party list. The vote stays with her until she is elected in count 259. At this point, all of the remaining Greens, and the Socialists, have been eliminated. The next highest ranked  continuing candidate is Fiona Patten, the first listed candidate of the LCP (preference 9 on the ballot). She received the ballot at a value of 0.0322 (0.968 of the ballot was used to elect Steph HODGINS-MAY). The ballot stays with Fiona PATTEN until she is eliminated in count 284. At this point, the next highest ranked  continuing candidate is Michelle ANANDA-RAJAH of the ALP (preference 19 on the ballot). The candidates preferenced at 10 to 18 had either been eliminated prior to count 284 (the remaining LCP and the AJP candidates) or elected (the ALP candidates ranked at 17 and 18). The ballot, now worth 0.0322 votes, helps Michelle ANANDA-RAJAH win the last  seat, at which point counting stops.

\section*{Challenges}

\begin{figure}[t]
    \centering
    \includegraphics[width=\textwidth]{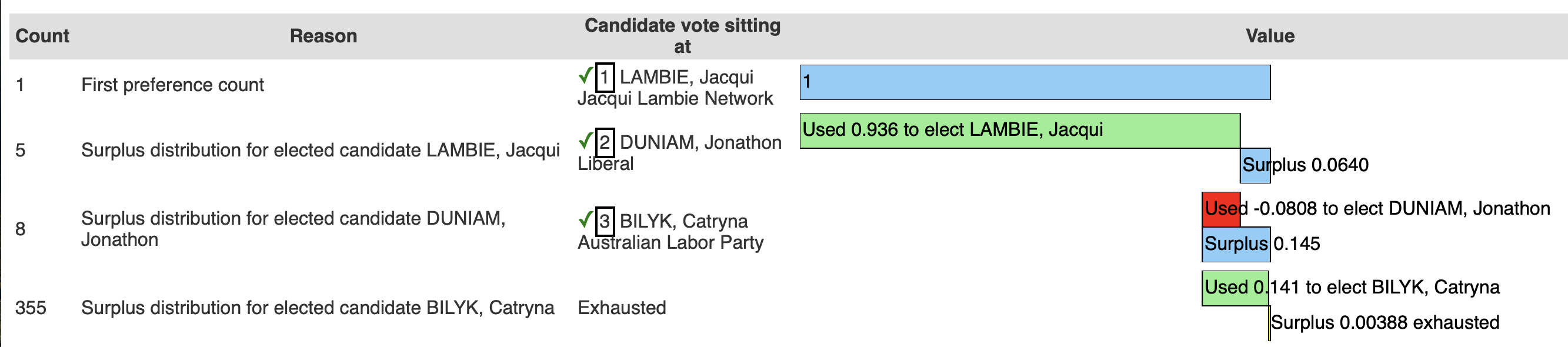}
    \caption{In the 2016 Federal Senate election for Tasmania, a vote that preferences Jacqui LAMBIE (first), Jonathon DUNIAM from the Liberal party (second) and Catryna BILYK from the Labor party (third) would actually have harmed Jonathon DUNIAM's chances of being elected due to contributing a negative amount to his tally.}
    \label{fig:TasNegative}
\end{figure}

Exploring the journeys of different votes surprised even veteran STV researchers like ourselves. This speaks to the complexity of the STV algorithm. At the bottom of the `To whom did my vote go?' page for each election is a section called `Challenges'. These show how specific votes in past elections led to some interesting situations. You may think that a vote could never contribute   a \textit{negative} amount to a candidates' tally. It was certainly possible in the 2016 Federal Senate election for Tasmania (\autoref{fig:TasNegative})!

\section*{Conclusion}

This paper describes a demonstration system for exploring vote journeys in past Australian STV elections. This system allows voters to see how specific types of votes, including those recommended by each party in their How to Vote cards, would have contributed to the tallies of relevant candidates over the course of tabulation. Systems like these may help voters better understand the nuances of Australian STV implementations.

\bibliographystyle{unsrtnat}
\renewcommand*{\bibfont}{\interlinepenalty 10000\relax}
\renewcommand{\bibsection}{\subsection*{References}}
\bibliography{references} 

\end{document}